\documentclass{article}
\newcommand{\be}{\begin{equation}}
\newcommand{\ee}{\end{equation}}

\begin{document}
\large
\hoffset=-3cm
\voffset=-1cm
\begin{center}
{\bf LARGE SCALE QUANTIZATION AND THE PUZZLING COSMOLOGICAL
PROBLEMS}\\ \vspace{0.8in} {\large Arbab
I.Arbab\footnote{arbab64@hotmail.com} \vspace{0.4in}
\\
Department of Physics, Faculty of Applied Sciences, Omdurman Ahlia 
University,
P.O. Box 786, Omdurman, SUDAN}
\end{center}
\vspace{2cm}
\begin{center}
Abstract
\end{center}
{\sf We have investigated the implications of Quantum Mechanics to
macroscopic scale. Evaporation of Black Holes and evolution of
pulsars may be one of the consequences of this conjecture. The two
equations $GM=Rc^2$ and $GM^2=\hbar c$ where $R$ and $M$ are the
radius and the mass of the universe, are governing the evolution
of the universe throughout its entire cosmic expansion, provided
that the appropriate Planck constant is chosen. The existence of
very large values of physical quantities are found to be due to
cosmic quantization. We predict a constant residual acceleration of the
order $10^{-7}\rm cm s^{-2}$ acting on all objects at the present
time.
\\
\\
PACS(numbers): 03.65, 98.80
\\
\\
\\
\\
{\bf \large 1. INTRODUCTION}
\\
  \\
The general theory of relativity (GTR), which attributes  the structure of 
space-time
to the gravitating mass, is incompatible with the theory of
quantum mechanics (QM). While quantum mechanics is a linear theory the GTR 
is highly
nonlinear so that the two theories are dissimilar. Attempts to linearize the
GTR do no give all manifestation of the theory. Initially the intention of 
the development
of GTR  was to describe the large scale structure of the universe, i.e., 
solar and
galactic scales, whereas QM deals with the microscopic scales.
The unification of electromagnetic, weak and strong was successful and gave
a lot of hints to further unification with the hitherto unachievable 
gravitational
interaction. Particle physicists are ambitious to get the unification of all 
interactions
at an energy scale of $10^{19}\rm GeV$!

However, these disparate worlds can meet in some cases. In these situations,
phenomena in atomic physics appear to be written in the language of gravity
and Einstein's GTR. By knowing the working of one subject, we can thus make
educated guesses about the other.\\
One therefore should not look for a paradigm in which gravity is manifested 
as a force,
but we should treat gravity  as a background (framework) on which other 
interactions occur.
Thus gravity provides only the shape of the space-time membrane on which 
other interactions
are carried.
We therefore, expect that gravitational effects are always present and 
manifested
in the way space-time affects our physical phenomena.
We remark that one should treat gravity not as an independent interaction 
but
rather as a framework in which all interactions take place and consider
only the strong, weak and electromagnetic interactions to be independent.
We therefore expect to observe quantum effects at microscopic as well as 
macroscopic
scale depending on the dimension of the system under consideration. This 
would imply
that quantum effect at cosmic scale to be similar to those at microscopic 
scale,
they only differ in the magnitude of the quantization. This is plausible 
since the masses
and distances are enormous for macroscopic system.\\
After making these arrangement we would end up with a cosmic Planck constant
having a huge value for macroscopic scales. The relative  errors in 
determining the
physical properties of the macroscopic system would have the same value as
for microscopic one. One would also expect that the same laws governing the 
microscopic scale
to be still functioning but with scaled ones. This might require some 
physical quantities
to evolve in order to satisfy these laws.

Based on Tiftt's [1] experimental data on galactic red-shift DerSarkissian 
[2]
suggested  a cosmic quantum mechanics (CQM) characterized by  the Planck
constant [3]
\be
h_{\rm g}=10^{102}h =10^{68}\rm\ J.s
\ee
where $h$ is the ordinary Planck constant of the
quantum mechanics. The quantum state of a galaxy is described by a wave
function $\psi(x,t)$ so that galaxies become the elementary particles of 
CQM.
For a gravitational system and according to Einstein equivalence principle
a body with observable  inertial mass ($m_i)$ has observable active mass 
($m_a$)
and hence an observable gravitational field. The  observability of the this 
field
requires that the total energy (self-gravitational energy) stored in it is 
measurable
and, therefore, greater than $h/2t$ (according to the uncertainty principle
$\Delta E\Delta t> h/2$ and $\Delta t\sim t, \Delta E\sim E$). For a 
spherical body
of mass $M$ and radius $R$ we have
\be
\frac{GM^2}{R}>h/2t.
\ee
But we know, from general relativity,
that $\frac{GM}{Rc^2}\le 1$ so that $ Mc^2> h/2t$.

In QM  a body is observable only if its self-gravitational energy $E_G$ is 
greater
than $h/2t (E_G>h/2t$). All bodies in QM obey this inequality.
By supposing that the CQM obey the same rules as QM except that the 
corresponding
Planck constant is different we can write,
\be
E_G>h_c/2t
\ee
which is obeyed by all cosmic bodies.
For galaxies this formula gives
\be
\frac{GM^2}{R}>h_c/2t.
\ee
This implies that,
for $t=10^{17}\rm sec$, $ R=\rm 10^{19}m$, and
\be
h_c=\rm 10^{68}J.s,
\ee
$M>\rm 10^{40}kg$.
This represents  a constraint for galaxies obeying CQM (however all galaxies 
obey this).
A system with spatial dimension $R$,  mass $M$, spin $S$ and a positive 
energy
density has $ R>\frac{S}{Mc}$ [4]. This together with eq.(4), yield the 
inequality
\be
M^3>\frac{Sh_c}{2Gct}
\ee
that  holds for cosmic bodies obeying CQM with intrinsic spin angular 
momentum.
The quantization of spin requires that $S=nh$ ($n$ is an integer). Hence,
\be
M^3>\frac{h^2_c}{Gct}
\ee
valid for spinning cosmic bodies. By considering a cluster
made  by many galaxies with quantized spin Ruffini and Bonazzola [5]
  have shown that there exists no equilibrium
configuration for a cluster of more that $N\sim (\frac{M_P}{M})^A$ galaxies
of mass $M$ in their ground state, where $M_P$ is the cosmic Planck mass
\be
M_P=(\frac{h_cc}{G})^{1/2}\sim \rm 10^{43}kg,
\ee
where $A=2$ for bosons, $A=3$ for fermions.
Thus CQM applies  well to galaxies. Is there other CQM that applies at 
planetary,
solar or universal scale with different quantum of action ($h_c$)?

Caldirola {\it et al} [6] suggested in a framework of unified theory of 
strong and
gravitational interaction that a quantum of action for the universe is given 
by

\be
h_c=MRc
\ee
where $M$, $R$ are the mass and the radius of the universe.
Equation (9) gives
\be h_c=\rm 10^{87}\ \ J.s
\ee
with $M=10^{53}\rm\ kg$ and $R=10^{26}\rm\ m$ [6].\\
A possible connection between $h$ and $h_c$ for the universe exists
where $\frac{h}{r^3}\sim\frac{h_u}{R^3}$ where $r$ is the size of a
typical hadron and $h_u$ regarded as nonvanishing total angular
momentum of the universe due to torsion field [7].
  Massa [3] enlarged the above relation to become
  \be
  \frac{h}{r^3}\sim\frac{h_u}{R^3}\sim\frac{h_g}{R_g^3}
  \ee
where $R_g$ is a typical radius of a galaxy ($R_g\sim \rm 10^{19}m)$.
It was suggested by Caldirola {\it et al} [6] that ``these forms are special 
case of a more general form still
unknown''. In fact, the CQM for planetary scale is introduced by Pierucci 
[8]
in which
\be
h_c=\rm 10^{-13} J.s
\ee
which explains the Titius-Bode law and other
regularities of the solar system.
Massa considered the possibility of the growth of $h_c$ with cosmic time 
($t$). He, according
to the Large Number hypothesis (LNH) [9], suggested that
for a flat universe $h_c\sim t^{5/2}$, and he concluded that the CQM is 
incompatible
with this LNH. Carneiro [10] investigated the extension of scale invariance 
to quantum
behavior. The price that paid was the scaling of Planck constant ($h$)
leading to quantization of large structure that treated till now as 
classical systems.
He used the LNH in his theory and concluded that the angular momentum for
a rotating universe is of the order of magnitude of $\rm 10^{87} J.s$.

So far there is yet no evidence that the universe is rotating but if it does 
it should
do that with this value! However, Kuhne [11] claimed an observation of a 
rotation
of the universe.
We remark that an order of magnitude for the universe angular momentum
is within the limit for the global rotation obtained from the  cosmic 
microwave
background radiation (CMBR) anisotropy obtained by Kogut {\it et al} [12].
Carneiro [10] suggested that ``one of the possible explanation of the large 
scale quantization
could be based on an evolutional point of view: the quantum nature of the 
universe
during its initial times has molded its-apparently quantized nowadays large 
scale
structure''.\\
We also  note that Nottale obtained, on a different ground, a scale Planck 
constant of
order $\rm 10^{42}J.s$ for a QM model for the solar system [13].\\
Carneiro [10] applied the Bohr quantization condition to the motion of Earth 
around the
Sun and found that this is consistent provided that Planck constant has the 
value
$\rm 10^{42}J.s$. Muradian [14] found an angular momentum of stars around 
$\rm 10^{42}J.s$ which
is close to Kerr limit of a rotating black hole with a mass of $\rm 
10^{30}kg$.
\\
In this work we try to explain the origin of large numbers
occurring in nature by appealing to this conjecture. The
hierarchical  nature of mass, cosmological constant, entropy,
temperature and angular momentum are found to be one of the
manifestation of this conjecture. The evolution and
characteristics of astrophysical objects like, black holes and
binary pulsars are investigated in the framework of this
conjecture. The huge quantum number is absorbed in the definition
of the cosmic Planck constant. These numbers are in the order of
$10^{20}, 10^{40}, 10^{60}. 10^{80}, 10^{100}, 10^{120}$. This
makes our universe appear to have a Planckian manifestation, i.e.
all of the present cosmological parameters are one of the above
numbers times the Planckian quantity. In this work the Large
Number Hypothesis finds its physical justification.

We present our
model in sect.2 and discuss its implication in a framework in
which $G$ and $\Lambda$ vary with cosmic time in sec.3. In sect.4
we apply our conjecture to Black Holes and Binary Pulsars. In
sec.4 we constrain the maximal acceleration and force appearing in
the universe, according to  our model. We wind up our paper with a
conclusion.
\\
\\
{\bf \large 2. THE MODEL}
\\
\\
The usual definition of Planck mass ($m_P$) can be inverted to give
\be
h=\frac{Gm_P^2}{c}.
\ee
In a similar fashion, we suggest a cosmic Planck constant ($h_c$) with a
cosmic Planck mass $M_P$ as
\be
h_c=\frac{GM_P^2}{c}.
\ee
Another possibility, from a dimensional point of view, is
\be
h_c=\frac{c^3}{G\Lambda},
\ee
where $\Lambda$ is the cosmological constant. This is equivalent to writing 
it
as $\Lambda=\frac{1}{L_P^2}$ where $L_P$ is the Planck length.
By supposing that the energy density of the vacuum is caused by the 
gravitational
interaction of the neighboring particles with mass $m$, Lima and Carvalho 
[15]
obtained
\be
\Lambda=\frac{G^2m^6}{\hbar^4}.
\ee
from which one can write
\be
h_c=\frac{G^{1/2}M_P^{3/2}}{\Lambda^{1/4}}.
\ee
Hence, we have three different forms of cosmic Planck constant to be tested 
with observation.
\\
\\
{\bf\large 2. A MODEL WITH VARIABLE $G$ AND $\Lambda$}
\\
\\
We have shown in an earlier work that [16] $G\sim t^2$ and $\Lambda\sim 
t^{-2},
m\sim t^{-1}$ in the early universe.\\
Therefore, all three forms reduce to the ordinary Planck constant in the 
early universe
(radiation dominated era).
Thus one has $h_c=h$ in the early universe.
Hence CQM and QM are equal in the early universe.
However, in the matter dominate universe [16] $G\sim t, \Lambda\sim t^{-2}$ 
and $m=\rm constant$.
Therefore $h_c\sim t$. This asserts that the Planck constant evolves with 
time and
that why it has different values for different systems.
Thus QM applies to macroscopic system in the same way as in microscopic 
system
except Planck constant is replaced by the cosmic Planck constant.

Equation (15) shows that the quantum of action for the universe is related 
to the vacuum
energy density ($\Lambda$) of the universe, so that one obtains
\be
\frac{\Lambda_P}{\Lambda_0}=\frac{h_c}{h}=10^{120}\ .
\ee
This gives a present value for $\Lambda$ (i.e., $\Lambda_0$) a value
of $\rm 10^{-52}m^{-2}$ as expected from present observations.
Equations (15) and (17)  yield
\be
\Lambda=\frac{c^4}{G^2M_P^2}\ \ ,
\ee
as a cosmological law governing the evolution of the universe. This implies
that in the early universe [16] ($G\sim t^2, m\sim t^{-1}$) $\Lambda\sim 
t^{-2}$.
Similarly in the matter dominated phase ($G\sim t, m=\rm const.)\ \  
\Lambda\sim t^{-2}$.
Hence $\Lambda\sim t^{-2}$ during  both phases. To reproduce the result in 
eqs.
(1), (5) and (10) one requires $M^u_P=\rm 10^{53} kg$,
$M^g_P=\rm 10^{43} kg$ and $M^s_P=\rm 10^{30} kg$, representing the 
corresponding
Planck masses at universal, galactic  and solar scales, respectively.
We remark that cosmic Planck constant for the planetary system is 
$10^{26}\rm kg$.
This represents the average mass of the planets in the solar system.
Hence,
\be
h_c=\frac{GM_P^2}{c}=10^{34}\rm J. s\ .
\ee
This value is different from the Planck constant for the planetary system
quoted in eq.(12).
Thus eqs.(14), (15) and (17) provide a bridge connecting macroscopic  and 
microscopic
phenomena through a simple formula.
It is found that the universe satisfies the equation
\be
GM=Rc^2\ .
\ee
This is an statement of the equality of the rest mass energy of the universe 
to its
gravitational energy (i.e., $Mc^2=\frac{GM^2}{R}$).
We observe that eq.(9) is same as eq.(14), if we use the Mach
relation (eq.(21)) [25].

We have shown in an earlier work [16] that eq.(21) gives
$R\sim t$,  so that eq.(21) yields
\be
M\le\frac{c^3t}{G}.
\ee
which is similar to eq.(7). This inequality is obtained by [17] by different
approaches.
One may define a maximal mass for a bound system, at time $t$, to be 
gravitational stable with
\be
M_{max.}=\frac{c^3t}{G}.
\ee
So that the above equation becomes
\be
M\le M_{max.}\ .
\ee
This formula is obtainable from Friedman cosmology ($3H^2=8\pi G\rho$) with
$R\sim t, H\sim t^{-1}$ giving $M=\frac{c^3t}{G}$.\\
If we consider the stars to be the atoms of the universe we will observe 
that
the universe is a typical one {\it solar mole}, consisting of $10^{23}$ 
stars(suns).
\\
The $M_{max.}$ defined  above would have a significance for the
time $t=1$ sec and $t=10^8$ sec. The former defines the Planck
mass for the stars and the latter for galaxies. Specifically,
\be
\rm M_{max.}=10^{35}kg \ \ \ and\ \  M_{max.}=10^{43}kg\ , \ee
respectively. Hence the stars and globular galaxies today were
once the actual states of our past states of our universe. That is
why one need not go back in time to see how was the past. Thus our
universe may contain information about the past with it at all
times and not just cosmic background radiation! For these system
the Planck constant evolves as
\be
\hbar_c=\frac{c^5}{G}t^2 \ee During this time the gravitational
constant has remained unchanged. Hence, stars and globular
galaxies would have the following cosmic Planck constants:
\be
\rm 10^{52}J.s \rm\ \ and\ \ 10^{68}J.s, \ee respectively.
However,these values are reported by Capozziello {\it et al.}[27].
\\
We now turn to calculate the Compton wavelength of the universe,
i.e., $\lambda_U=\frac{h_c}{Mc}=\rm 10^{26}m$, which is of the
same order of magnitude of the present radius of the universe.
Thus the use of CQM for the present universe is also logical and
plausible. Thus all known bounded systems are characterized by
their cosmic Planck constant.\\ We would like to remark that the
magnetic dipole moment of the Earth (E) is found to be $
7.98\times 10^{25}G\ cm^3$. This value can be obtained from the
definition of the magnetic dipole moment, i.e.,
$\mu_E=\frac{e\hbar}{2mc}$ for an atomic system. Here we make the
following replacement: $e^2=Gm^2$ and $\hbar=10^{34}\rm J.s$,
$m=M=6\times 10^{24}\rm kg$. This gives the same order of
magnitude for the magnetic dipole moment of the Earth. Thus, as we
remarked in the beginning, having known the atomic system
parameters their gravitational analog can be obtained by the
generalization above. We also note that $\mu\propto m^2$. Hence,
one can conclude that for Mercury (M) this gives
$\rm\mu_M=\mu_E(\frac{m_M}{m_E})^2=7.98\times10^{25}(0.0553)^2=2.4\times10^{22}\
G\ cm^3$. This in fact is the presently known value for Mercury.
We observe that there is a new large numbers associated with our
physical world. These numbers are $(10^{34})^a$, where $a =-1,
2,2.5, 2.75, 3, 3.5$. These {\it magic} numbers identify bound
structures that are present in our universe today; the atoms,
planets, stars (suns), globular galaxies, galaxies and the whole
universe. Thus these structures represent self-similar systems.
And all these systems manifest the quantum behavior, but in
different ways. The above quantum numbers can be compared with
Dirac large numbers $(10^{40})^n$, with some $n$. Here we have
seen that the cosmic Planck constant is a multiple of $10^{68},
10^{85}, 10^{102}, 10^{119}$ times the ordinary Planck constant.
Or, equivalently, the quantum numbers for large-scale system are
$10^{68}, 10^{85}, 10^{102}, 10^{119}$, instead of the $1, 2, 3,
..$ for the microscopic system.
\\
\\
{\bf\large 3. APPLICATION OF CQM TO BLACK HOLES}
\\
\\
Consider a spinning ($S$) black hole (or neutron star) with frequency 
$\omega$.
We have
\be
S=I\omega, \ee with $I=MR^2$, $M$ and $R$ are the mass and radius
of the object. For an object with a gravitational radius
$R=\frac{2GM}{c^2}$ and spin $S\sim h_c$, eq.(28) yields
\be
\omega=\frac{c^3}{GM},
\ee
Pulsars are believed to be rotating neutron stars and a newborn pulsar 
formed
in supernova may be rotating with a frequency of $\rm 10^3s^{-1}$
emitting a gravitational radiation with a rate of $\rm 10^{48}Js^{-1}.$
For a pulsar of a mass $M=M_\odot$
one gets, from CQM, a period of $\rm 10^{-3}$ sec.
One of the most promising source is the pulsar NP0532 in the Crab nebula 
[18]. This pulsar is
observed to emit pulses of electromagnetic radiation, at optical, X-ray and 
radio
frequencies with a period of 33 $\rm m sec$. Thus our model, though based on
rough estimates, is in a good agreement with observations.
\\
A black hole emitting radiation as a black body with a temperature $T$
given by
\be
k_BT=\hbar\omega. \ee Using eqs.(29) one gets
\be
T=\frac{\hbar c^3}{GMk_B}
\ee
in comparison with Hawking [19] formula obtained from QM treatment for a 
non-rotating black hole,
viz.,
\be
T=\frac{\hbar c^3}{4\pi GMk_B}.
\ee
A rotating galaxy of mass $\rm 10^{43}kg$ would have a frequency of $\rm 
10^{-8}s^{-1}$ or
a period of $10\rm\ years$.
We can compare this value with the presently observed rotation of galaxies.
The time for the evaporation of the black hole can be estimated from the 
uncertainty relation
\be
\Delta E\Delta t\sim h_c
\ee
with $\Delta E=Mc^2$ and $\Delta t=\tau$.
This upon using eq.(14) becomes
\be
\tau=\frac{GM}{c^3}. \ee Since $\frac{GM^2}{c\hbar}=1$, one can
write eq.(34) as
\be
\tau=\frac{GM}{c^3}.\frac{GM^2}{c\hbar}=\frac{G^2M^3}{c^4\hbar}\ .
\ee
Which is obtained by Hawking [19] from a quantum mechanical treatment.
Hence, we may write for the evaporation of Black holes the formula
\be
\tau=\frac{G^2M^3}{c^4h_c} \ee as a CQM analogue.\\ We see that a
black hole of one Planck mass $m_P$ evaporate during Planck time
($\rm 10^{-43}s$). A galactic rotating black mass evaporates
during a time of 1 year while a solar rotating black hole
evaporates during a time of $\mu\ sec$. A rotating black hole of
size of the universe evaporates during  a time of $10^{10}$ years.
\\
The entropy of a black hole is given by [19]
\be
S=\frac{GM^2}{c\hbar}k_B
\ee
which upon using eq.(14) yields
\be
S=\frac{\hbar_c}{\hbar}k_B=(\frac{k_B}{\hbar}){\hbar_c}
\ee
This entropy is independent of the mass of the object in question as long as
$\hbar_c$ describes that object. We see that for a black hole
formed in the early universe $\hbar_c=\hbar$, and therefore irrespective of 
its mass
the black hole will have one unit of entropy, i.e., $S=k_B$.
Black holes forming during solar and galactic time will have entropy
that is multiple of $\hbar_c$. We thus conclude that the entropy of black 
holes
is quantized. A galactic mass black hole  will have an entropy of $\rm 
10^{102}k_B$,
while a solar mass black hole will have an entropy of $10^{76}\rm k_B$.
However, a black hole of the mass of the universe has an entropy of 
$10^{120}\rm k_B$.
\\
\\
{\bf \large  4. THE VACUUM ENERGY DENSITY\footnote{see ref.[26]
for a review about vacuum energy}}
\\
\\
The Planck energy density is defined as
\be
\rho_P=\frac{c^5}{G^2\hbar} \ee which represents the maximum
density of the universe at Planck time. We now employ the CQM and
evaluate the present maximum energy density of the universe, i.e.,
\be
\rho_P^0=5.4\times \rm 10^{-28}gcm^{-3}. \ee Present observations
set a limit on the present energy density as
\be
\rm 10^{-30}gcm^{-3}<\rho^0<10^{-29}gcm^{-3}. \ee We have thus
obtained a constraint on the density of the universe viz.,
$\rho^0\le \rho^0_P$. It is evident that the Planck energy density
of the universe indeed evolves with time.\\ We argue that the
Planckian energy density ($\rho_P$) is equal to the vacuum energy
density of the universe at all times. In this sense the universe
is still governed by quantum mechanics (cosmic). We have, from
Einstein de Sitter model, the relation
\be
3H^2=8\pi G\rho \ee Upon using eq.(39) this becomes
\be
GH^2\hbar=\frac{8\pi}{3}c^5=2\times 10^{43} \ee This equation may
be taken to define a cosmic Planck constant. Thus
\be
\hbar_c=\frac{2\times 10^{43}}{GH^2}\ . \ee In the early universe
[16] we have $G\propto t^2$, $R\propto t$, so that the cosmic
Planck constant coincides with the ordinary Planck constant
$\hbar$. At the present time, we see that the cosmic Planck
constant, $h_c=10^{87}\rm J.s $. In the present universe we have
[16] $G\propto t$, $R\propto t$, so that eq.(44) implies
$\hbar_c\propto t$. This confirms our earlier findings of cosmic
Planck constant for the whole universe. Therefore (44) represents
an additional definition of cosmic Planck constant to those
already found in eqs.(14), (15), and (17).

The vacuum energy density is defined as
\be
\rho_v=\frac{\Lambda c^2}{8\pi G}=\frac{c^5}{G^2\hbar} \ee During
the radiation dominated phase this yields
\be
\rho_v=\rho_{vP}(\frac{t_P}{t})^4, \ee where $\rho_{vP}$ is the
vacuum energy density at Planck time. This shows that $\rho_v$
behaves like radiation in the early universe. During the matter
dominated phase one gets
\be
\rho_v=\rho_{v0}(\frac{t_0}{t})^3, \ee where $\rho_{v0}$ is the
vacuum energy density at present time. Similarly $\rho_v$ scales
like matter in the matter dominated phase. This would mean that
the vacuum contribution to the energy density of the university
has been very significant in the past and in the present time. It
is therefore not surprising to find out that the vacuum contribute
$66\%$ to the total energy density of the present universe. Thus
one may resolve the dark matter energy that is postulated to
circumvent some astrophysical problems. We remark that since the
vacuum energy density was equal to the Planck energy density in
the early universe it still governed by this quantum energy. This
is plausible since the Planck energy density energy with time. One
thus could say that our universe today is a typical quantum
system. Moreover, the Planck length has become today our physical
radius of the universe. We also see that the Planck length
$L_P=\sqrt{G\hbar/c^3}$ has the same time evolution in both the
early phase and the present matter  phase, viz.
\be
L_P(t)=L_P(\frac{t_P}{t})
\ee
and
\be
L_P(t)=L_P(\frac{t_0}{t})
\ee
Equation shows that the vacuum's quanta shifts like photon according to the 
Planck
law. Thus if they had different background temperature the former would cool
(at the present time and in the early universe) as
\be
T_v(t)=T_{vP}(\frac{t_P}{t}) \ee where $T_{vP}$ is the vacuum
temperature during Planck time (see eq.(57)). We observe that in
the early universe we have
\be
m\propto t^{-1} \ \ \ {\rm and}\ \ \ T\propto t^{-1}\propto R^{-1}
\ee
so that if these relations hold throughout the cosmic expansion, one would 
obtain
the relation
\be
m=m_P(\frac{t_P}{t})
\ee
where $m_P$ is the Planck mass at Planck time and $T$ is the temperature.
We note that De Sabbata and Sivaram [20]
relate the temperature ($T$) to curvature ($\kappa$) and showed that 
$T\propto \sqrt{\kappa}$,
but the time ($t$) scales as $t\propto \frac{1}{\sqrt{\kappa}}$. For a 
maximal
curvature $\kappa_{max.}=\frac{c^3}{\hbar G}$, which implies
\be
G\propto t^2, \ \ \ \rm and\ \ \ T\propto t^{-1}.
\ee
Comparison with eq.(15) immediately yields
\be
\kappa_{max.}=\Lambda\ . \ee Thus one may connect the smallness of
the present value of the cosmological constant to the flatness of
our present universe.\\ Using eq.(52) one would obtain
\be
m=10^{-5}(\frac{10^{-43}}{10^{18}})=10^{-66}\rm g. \ee which
represents  a minimal mass scale at the present time. Equation
(50) gives
\be
T_{v}=10^{32}(\frac{10^{-43}}{10^{18}})=10^{32}\times
10^{-61}=10^{-29}\rm K, \ee where $T_P=10^{32}\rm K$. We remark
that De Sabbata and Sivaram  obtained a similar value by
considering a time-temperature uncertainty relation ($\Delta
t\Delta T=\hbar/k_B$) and relate the maximum time to Hubble time.
They suggested that they could obtain such a value by considering
a black hole of a mass of the universe using the formula outlined
in eq.(31). Or by considering the maximal possible entropy of
$10^{120}\rm k_B$ which would imply this minimal temperature. They
also found a similar value and noted that this minimal temperature
corresponds to the quantum fluctuations of cosmological torsion
background.\\ Massa [21] has obtain a similar value and relates
this to the mass of graviton. He argued that in an expanding
universe this mass depends on time. Rosati [22] found the quantum
field today has typically a mass of the order of $10^{-66}\rm g$.
Larionov [23] attributed a similar mass term to an effective mass
associated with the vacuum energy density (or $\Lambda$). He
assigned this extremely low value of this effective mass to a
quantum with wavelength equal to the present radius of the
universe. One may also add to this conjecture the cosmological
constant problem (rooted in the enormous value, i.e.,
$\Lambda_P=10^{120}\Lambda_0$) as due to the cosmic quantization,
as is evident from eq.(18)!In fact, Zizzi [28] has brought an idea
of quantized cosmological constant that is in the same line as our
reasoning here.
\\ According to Massa assertion the graviton has a mass; this
would mean that the range of the gravitational interaction is not
infinite  but limited by this mass scale. Hence, the maximum
possible interaction distance between any two gravitating objects
has to be at a maximum distance of $10^{26}\rm m$. A similar
assertion would also hold for electromagnetic interaction if it
turned out that a photon is not massless!
\\
We have so far shown that the two formulae (eqs.(14) and (21))
\be
GM=c^2R\ \ \ \rm and\ \ GM^2=\hbar_cc
\ee
hold throughout the cosmic evolution of the universe, provided we consider 
the CQM
to be a valid principle.
\\
\\
{\bf \large  4. MAXIMAL ACCELERATION AND FORCE}
\\
\\
The self-gravitational force of a system of mass $M$ and radius $R$ is given 
by
\be
F=\frac{GM^2}{R^2}. \ee Using eq.(57)
\be
F=\frac{c^4}{G}\ ,
\ee
for the universe. This force is clearly independent of the mass of the 
object into
consideration. It is thus a universal force, and since it depends on $G$ 
inversely
it defines a maximal self-gravitational force. There corresponds to this
maximal force a maximal acceleration ($F_{max.}=Ma_{max.}$) defined by
\be
a_{max}=\frac{c^4}{GM}.
\ee
If we consider a variable gravitational constant as suggested in [16], one 
gets a minimal
gravitational force in the universe during the nuclear (or hadronic) epoch 
as
\be
F_{min.}=10^{-2}N\ ,
\ee
since during nuclear (or hadronic) epoch the gravitational constant was
$G_N=10^{40}G_0$.
The smallness of this force may account for the fact that quarks are
asymptotically free inside hadrons, according to the theory of quantum 
chromodynamics (QCD).
A maximal force in the universe during the present or Planck time is
\be
F_{max.}=10^{44}N\ .
\ee
We remark that the factor $\frac{c^4}{8\pi G}$ appearing in the Einstein's 
equation
may be interpreted as the force per area required to give space-time unit
curvature, that is $10^{43}Nm^{-2}$ for a curvature of $1m^{-2}$. Space-time
is therefore an extremely stiff medium.
De Sabbata and Siviram noted that there exists a maximal acceleration
given by
\be
a_{max.}=\frac{c^{7/2}}{G^{1/2}\hbar^{1/2}}, \ee originated as
quantum effect due to torsion. Their formula reduces to our
formula upon using eq.(14) into (60). Therefore, we have,
according to CQM
\be
a_{max.}=\frac{c^{7/2}}{G^{1/2}h_c^{1/2}}\ .
\ee
We now turn to calculate the maximal acceleration at a universal scale, 
according
to CQM.
\be
a_{max.}=10^{-9}ms^{-2}\ . \ee However, De Sabbata obtained a
value of $\rm 10^{-10}ms^{-2}$ on different grounds that agrees
with [24].  Note that the maximal acceleration at Planck time was
$a_{max}=\rm 10^{51}ms^{-2}$. Very recently, Anderson {\it et
al.}[29] have found an unmodelled acceleration towards the Sun of
$\rm 8.09\times 10^{-8}cm\ s^{-2}$ for Pioneer 10 and Pioneer 11.
Kuroda and Moi [30] have measured the relative acceleration in
free fall of pairs of test bodies
(aluminium/copper-aluminium/carbon))and found it to be of the
order of $3\times 10^9\rm ms^{-2}$. One may therefore attribute
this acceleration to some residual quantum background filling the
whole universe having a Planckian temperature of the order
$10^{-29}\rm K$, as shown before.
\\
\\
\\{\bf\large 5. CONCLUSION}
\\
\\
We have extended the implication of quantum mechanics form
microscopic scale to include macroscopic scale. This extension
resulted in a lot of interesting consequences concerning the
evolution and characteristics of black holes and binary pulsars.
We have found that quantities like entropy, cosmological constant
and time are quantized for macroscopic scales. Limiting values for
temperature, entropy, angular momentum ($\hbar_c$), force,
acceleration are obtained due to this conjecture. The universe is
found to have an energy density less that Planckian density at the
present time. We have also shown that we have different Planckian
parameters for the universe for different time. The relation
$GM^2=\hbar c$ which was found to apply in the early universe is
still valid during other phases, provided that we replace the
ordinary Planck constant with the cosmic Planck constant, whose
value depends on the properties of the macroscopic entity. We have
provided a physical justification for the Large Number Hypothesis
advocated by Dirac. We have made a correspondence between atomic
and gravitational systems and that all atomic parameters have
their gravitational analogue. The comparison between the atomic
parameters and the large scale parameters always involves numbers
of orders of $10^{20}, 10^{40}, 10^{60}, 10^{80}, 10^{100},
10^{120}$ times the atomic parameters.
\\
%\\
%{\bf\large ACKNOWLEDGEMENTS}
%\\
%\\
%I would like to thank the Sudanese Physicists Association (SPA) for 
providing
%financial support and the Omdurman Ahlia University for a research grant.
%I would also like to thank Drs. H.M. Widatallah, O.I. Eid and
%O.F.Osman for their enlightening discussion.
\newpage
{\bf\large REFERENCES}
\\
\\
1. W. Tifft, {\it Astrophys. Journal} 206,(1976)38, {\it Astrophys. Journal} 
221(1978)756,
   {\it Astrophys.Journal} 236(1978)70, {\it Astrophys. Journal} 
257(1982)442, {\it Astrophys. Journal} 262(1982)44\\
2. M. DerSarkissian, {\it Lett. Nouvo Cimento} 40(1984)390,
   {\it Lett. Nouvo Cimento} 43(1985) 274\\
3. C. Massa, {\it Lett. Nouvo Cimento} 44(1985) 671\\
4. L. Motz, {\it Lett. Nouvo. Cimento} 26(1962)673\\
5. R. Ruffini and S.Bonazzola, {\it Phys. Rev.} 187(1969)1767\\
6. P. Caldirola, M.Pavsic and E.Recami, {\it Nouvo Cimento} B48(1978)205,
    A.S. Goldhaber and M.M. Nieto, {\it Phy. Rev.} 9(1974) 1119\\
7. V. de Sabbata and M.Gasperini, {\it Lett. Nouvo. Cimento}
25(1979)489\\ 8. M. Pierucci, {\it Nouvo Cimento} B21(1974)69\\ 9.
R.H. Dicke, {\it Nature} 192(1962)440\\ 10. S. Carneiro, {\it
Foundation of Physics Letters}11(1998)95\\ 11. R.W. Kuhne, {\it
astro-ph/9708109}\\ 12. A. Kogut, G.Hinshaw, A.J. Banday, {\it
Phys. Rev.} D55(1997)1901\\ 13. L. Nottale, {\it Fractal
Space-time and Microphysics: Towards a theory of
   scale relativity} (World Scientific, Singapore (1993) pp.311-321\\
14. R.Muradian, {\it Astrophys. Space Sci.}69 (1980)339\\ 15.
J.A.S. Lima and J.C. Carvalho, {\it Gen. Rel. Gravit.} 26(1994)
909\\ 16. A.I. Arbab, {\it astro-ph/9911311}\\ 17. H. Genreith,
{\it gr-qc/9909009, astro-ph/9905317}\\ 18. J. Weber, {\it Phy.
Rev. Lett.} 21 (1968) 395\\ 19. S.W. Hawking, {\it Commun. Math.
Phys.} 43 (1975) 199\\ 20. V. De Sabbata and Z. Zhang, {\it Black
Hole Physics}, Kluwer Academic Publisher (Netherlands) 1992.\\ 21.
C. Massa, {\it Lett. Nouvo Cimento} 44 (1985) 609, {\it Lett.
Nouvo Cimento} 44 (1985) 694\\ 22. F. Rosati, {\it
hep-ph/9908518}\\ 23. M. G. Larionov, {\it Astrophys. Space Sci.}
252 (1997) 139\\ 24. D. Lindley, {\it Nature} 359 (1992), M.
Milgrom and R.H. Sanders, {\it Nature} 362 (1992) 25\\ 25.
G.C.McVittie, {\it Cosmological Theory}, (London, 1950), D.W.
Sciama, {\it Modern Cosmology} (Cambridge,1971)\\ 26. M. Robert,
{\it gr-qc/0012062}\\ 27. Capozziello {\it et al, gr-qc/9901042}\\
28. P. A. Zizzi,{\it gr-qc/0008049}\\ 29. John D. Anderson {\it et
al}, {\it gr-qc/9808081, gr- qc/ 0104064}.\\ 30. K. Kuroda and
N.Moi, {\it Phys. Rev. Lett.} 62(1989) 1941
\end{document}